\documentclass[a4paper, 10 pt, conference]{ieeeconf}  
\usepackage{graphicx}
\usepackage{tikz,graphics,color,float,epsf}
\usepackage{authblk}

\IEEEoverridecommandlockouts                              
\title{\LARGE \bf
An Immersive Virtual Environment for Collaborative Geovisualization
}

\author{Milan Dole\v{z}al}
\author{Ji\v{r}\'{i} Chmel\'{i}k}
\author{Fotis Liarokapis}
\affil{Human--Computer Interaction Lab \protect\\ Masaryk University, Faculty of Informatics, Czech Republic \authorcr Email: {\tt 396306@mail.muni.cz, jchmelik@mail.muni.cz, liarokap@mail.muni.cz}\vspace{1.5ex}}

\begin{document}

\maketitle
\thispagestyle{empty}
\pagestyle{empty}

\begin{abstract}
This paper presents an immersive  virtual reality environment that can be used to develop collaborative educational applications. Multiple users can collaborate within the virtual shared space and communicate with each other through voice. To asses the feasibility of the collaborative environment a novel case-study concerned the education of a geography was developed and evaluated. The geovisualization experiment scenario explores the possibility of learning geography in a collaborative virtual environment. A user-study with 30 participants was performed. Participants evaluated and commented on the usability and interaction methods used within the virtual environment.
\end{abstract}

\section{INTRODUCTION}
Multi-user virtual environments are places, where users can meet and collaborate, play or just relax. It is commonly used in entertainment. From MUDs (multi-user dungeons) with a few players to MMOs (massive multiplayer online) with thousands of active players~\cite{Feng2007}, multi-user non-immersive virtual environments aren't a new thing. Now, first commercial digital games with multiplayer in immersive virtual reality (or VR for short) are appearing.

In 2016, commercially available headsets for immersive VR from multiple manufacturers emerged on the market, which could mean a next step in the evolution of multi-user virtual worlds. Among other things, VR is also getting cheaper and even smartphones start to support VR in some form. This helps significantly with the popularization of VR.

Another area that explores possibilities of VR is education. Lifeliqe\footnote{https://www.lifeliqe.com/} for example is a store of hundreds of study materials for elementary schools possible to examine not only in VR, but also in the augmented reality (AR). However it currently supports only single user. Multi-user education in VR could open new possibilities of distance education~\cite{redfern_naughton_2002}.

The rest of the paper is structured as follows. Section II describes related work focused mainly on the collaboration in the virtual environments from different point of views (communication, interaction or games to name a few). Section III describes the implementation methods used to develop the environment, network features and controls. Section IV explains the design of the task used in the experiment. It also contains a brief description of participants, who took part in the experiment. In the fifth section are shown observations from the testing and feedback provided by participants. The last section provides conclusions and future work.


\section{RELATED WORK}
Collaborative Virtual Environments (CVEs) can be understood as virtual worlds shared by users through a computer network~\cite{Benford2001}. CVEs have different application domains ranging from health-care (\cite{McCloyRobert2001},~\cite{Rizzo2011}) to education (\cite{Faiola2013},~\cite {Nikiforos2013},~\cite{Pan200620},~\cite{redfern_naughton_2002}).


Users are able to interact with objects within the CVE. They are able to see other users interacting thanks to the avatar visualization. We used simplified avatars in our system, but more complex avatars, possibly with incorporated eye-gaze~\cite{Garau2003} using eye-tracking technologies to provide stronger sense of immersion, are considered for further research.


For the field of education VR can play a huge role. 
In the research of Papachristos et al. for example is explored use of the virtual world in education with the game Second Life as a virtual classroom~\cite{Nikiforos2013}. Their study discovered ``students' experiences from the educational activities and their attitudes toward the virtual environment were positive and not affected by the design of the educational setting.'' This is a cue to explore the virtual education even further using the immersive head-mounted displays for VR.

In the past, collaborative virtual environments lacked the non-verbal communication cues~\cite{redfern_naughton_2002}. Thanks to the user's avatar visualization however, it is possible to communicate non-verbally to some extent. During the design of the CVE system, forms of a non-verbal communication were considered and, if possible, incorporated.



\begin{figure}[!htbp]
\centering
\includegraphics[clip,width=1.42in]{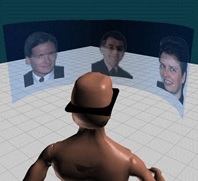}
\centering
\includegraphics[clip,width=1.8in]{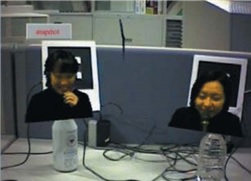}
\caption{Mixed reality~\cite{billinghurst_kato1999} (left) and augmented reality~\cite{Billinghurst2002} (right) videoconferencing.}
\label{fig:ar-videoconferencing}
\end{figure}

Collaboration is also an area of interest of Billinghurst et al., who experimented with collaborative mixed reality~\cite{billinghurst_kato1999} and collaborative augmented reality~\cite{Billinghurst2002}. In both cases they experimented with different techniques of computer supported collaborative work with the face-to-face communication (Fig. \ref{fig:ar-videoconferencing}). Face-to-face communication is however not possible in VR. The user wears a head-mounted display and his face is covered. The face-to-face communication thus has to be substituted with avatars.

Important problem to consider in the immersive VR is so called motion sickness. Potel described the motion sicknesses as follows: ``The most broadly accepted theory, called sensory conflict theory or cue conflict theory, holds that inconsistent sensory information about body orientation and motion causes the ill effects. Motion is detected by the semi-circular vestibular canals of the inner ear, which measure tilt and acceleration in six directions (roll, pitch, and yaw each in two directions). Body orientation is usually detected visually, but also by the internal muscular sensation of gravity's pull on the body.''~\cite{Potel1998} 

The problem with motion sickness should be considered during the design of any VR application. Currently known solutions, which help to avoid motion sickness are for example teleportation~\cite{Bozgeyikli2016}, tunneling VR locomotion technique~\cite{tambovtsev_floksy_peshe_2016} or walking--in--place paradigm~\cite{Tregillus2016}. In case of the teleportation and tunneling techniques, the user gets a feel that he's staying in one place. The walk--in--place paradigm requires additional motion of the body, that invokes the feel of walking.



\section{COLLABORATIVE VIRTUAL ENVIRONMENT}
\begin{figure}[!htbp]
\centering
\includegraphics[clip,width=3.3in]{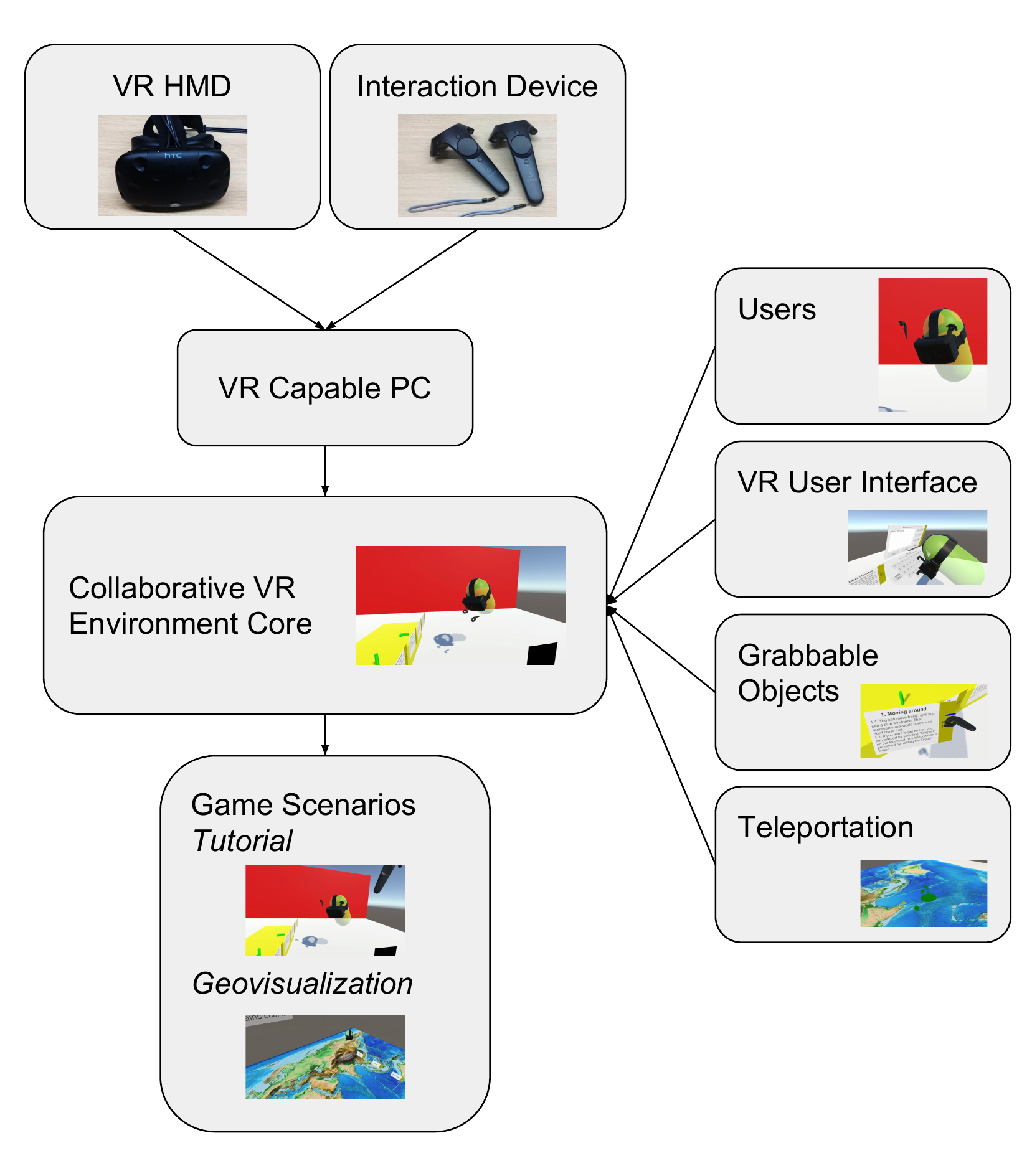}
\caption{System design overview.}
\label{fig:cvr-design}
\end{figure}

We developed a generic immersive virtual environment for multiple users. In this section we describe the implementation of the system and its components. In general, the core allows multiple users with VR HMD to share the same virtual environment. Within that environment, they can interact with objects and the user interface and these interactions are synchronized over the network. They can also move around by walking within the calibrated area or by using the teleportation to reach further distances. The users can see each other in the form of a simplified avatar. The overview of the whole system design can be seen on the Fig. \ref{fig:cvr-design}.

\begin{figure}[!htbp]
\centering
\includegraphics[clip,width=3.3in]{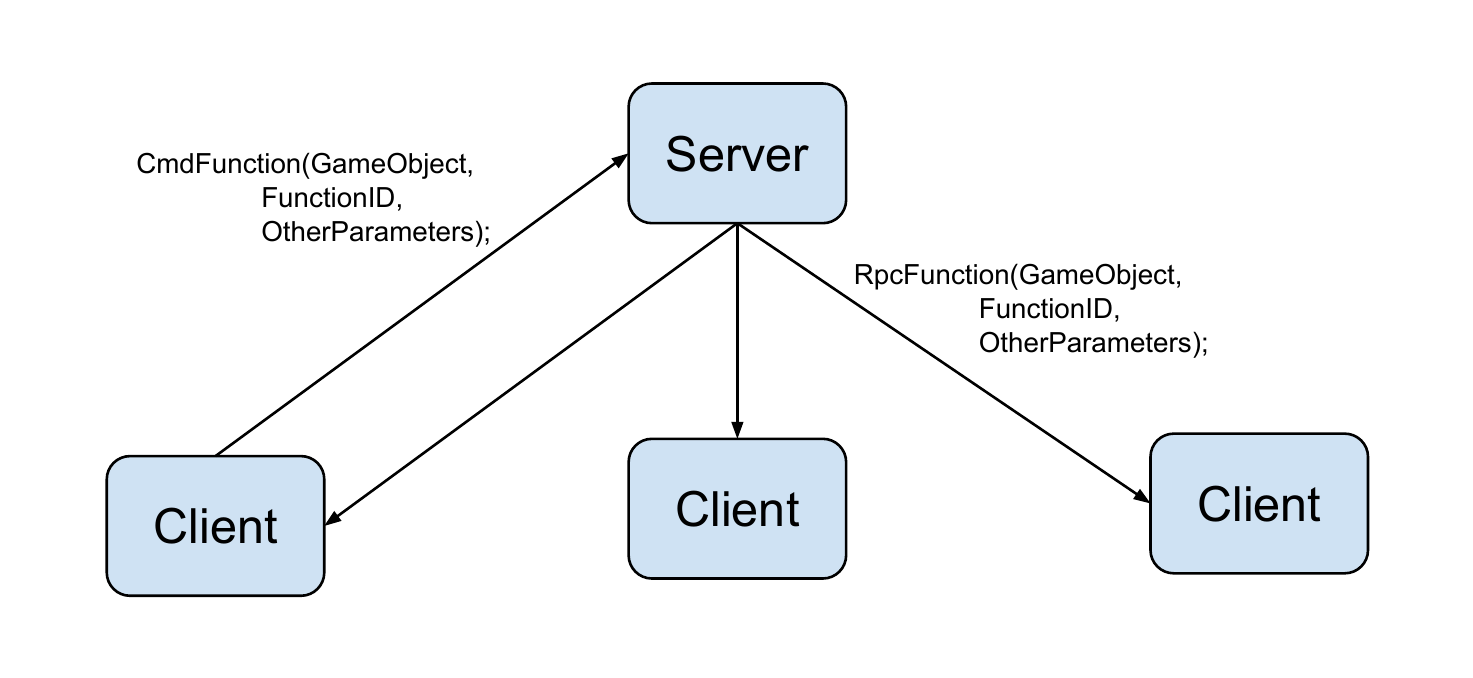}
\caption{Function calls on non-player objects.}
\label{fig:npo-communication}
\end{figure}

The core of the system is based on the Unity game engine with the use of SteamVR Toolkit. It uses a client-server model for networking. Communication between clients is achieved by sending commands to the server, which then sends Remote Procedure Call (RPC) messages to all clients (Fig. \ref{fig:npo-communication}). Such messages could be for example to change the rendering from the satellite to political map and vice versa.

On top of the core, we designed a case-study examining use of the system for geovisualization.


\begin{figure}[!htbp]
\centering
\includegraphics[clip,width=2.5in]{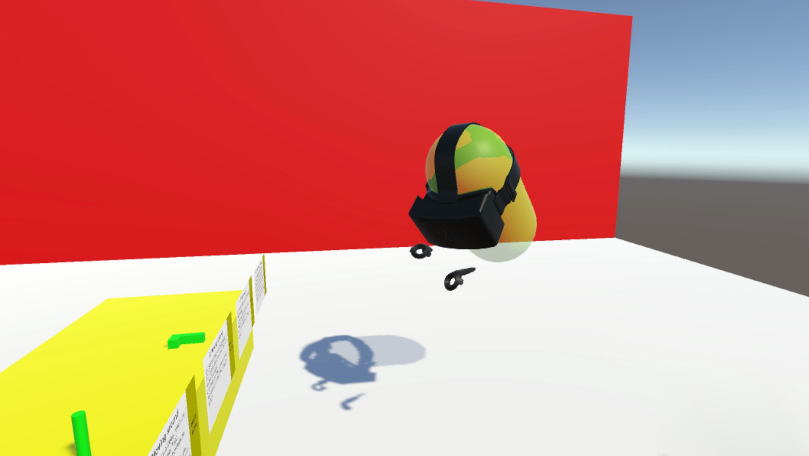}
\caption{Other user's avatar within the same virtual environment.}
\label{fig:cve}
\end{figure}

To allow users to see each other, it was necessary to add an avatar. Initially, the player was visualized simply as a capsule in space with floating controllers. This was enough to determine the user's location, but not the direction they are looking at. For that reason, model of the head-mounted display was added. Therefore other users were able to determine direction the user is looking at (Fig. \ref{fig:cve}).

Our system supports two types of users. One user has higher privileges above the others. That allows him to see objects that are hidden to others and thus control the environment. The role with higher privileges could be used by the teacher and the latter role by students.

\begin{figure}[!htbp]
\centering
\includegraphics[trim={28cm 5cm 0 14cm},clip,width=2.5in]{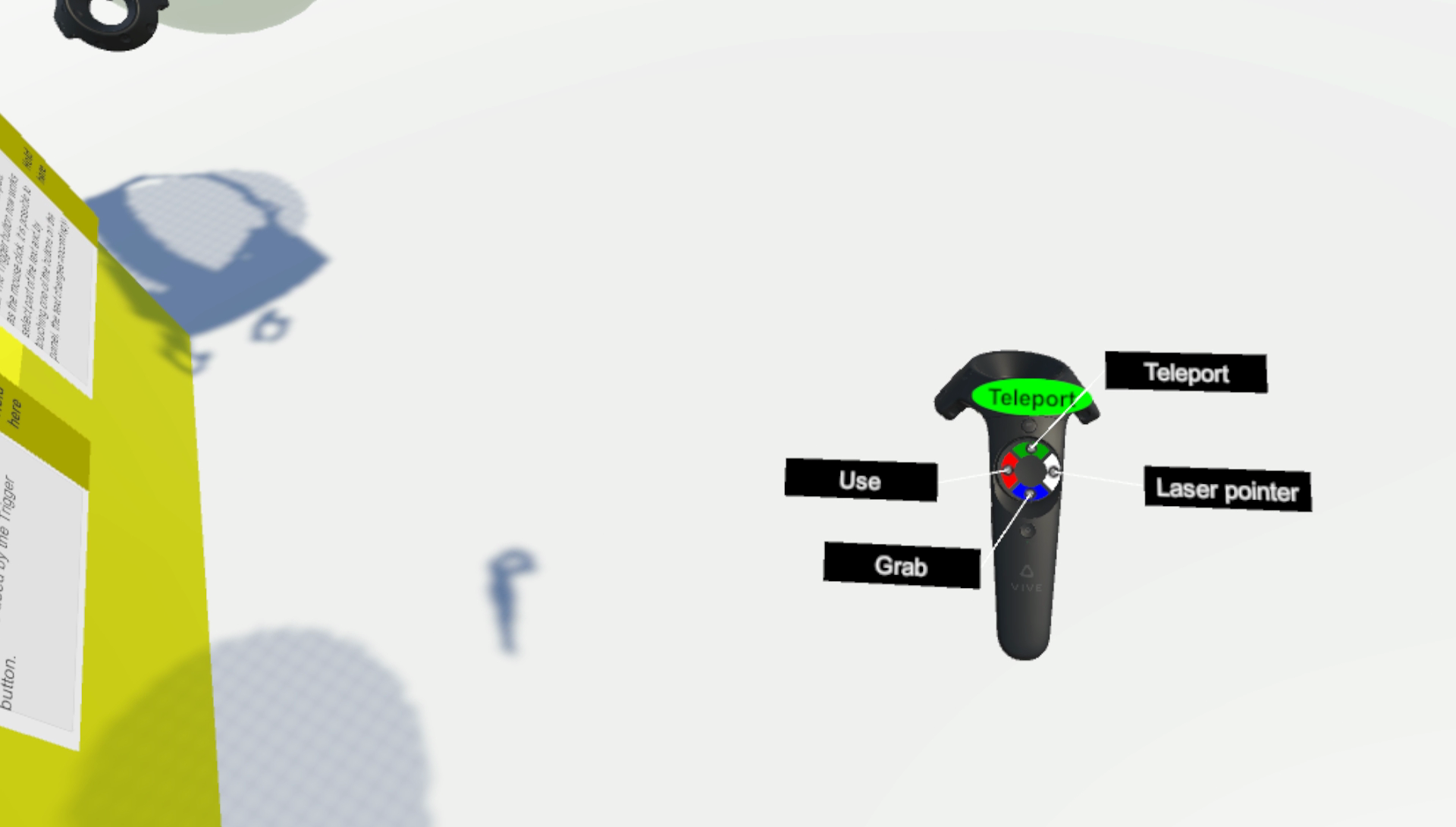}
\caption{Controller interaction selector with teleport currently selected.}
\label{fig:controller-ui}
\end{figure}

Emphasis was put into the simplicity of the interaction. The aim was to make it generic for later use in further experiments, yet intuitive enough so the explanation of the controls doesn't take take too much time. The controls was designed for the controller of the HTC Vive headset. Due to the limited number of buttons and bad perception of the ``grip'' button on the side of the controller, following controls was implemented. Touchpad is used as a selector for actions performed by the ``trigger'' button. Commonly used actions are teleportation for reaching further distance and object grabbing. Additional actions, like using held objects or laser pointer were implemented to add for future experiments (Fig. \ref{fig:controller-ui}).



\section{METHODOLOGY}
This section contains description of the design of the geovisualization scenario with an overview of the experiment procedure and its participants.

\begin{figure}[!htbp]
\centering
\includegraphics[clip,width=2.5in]{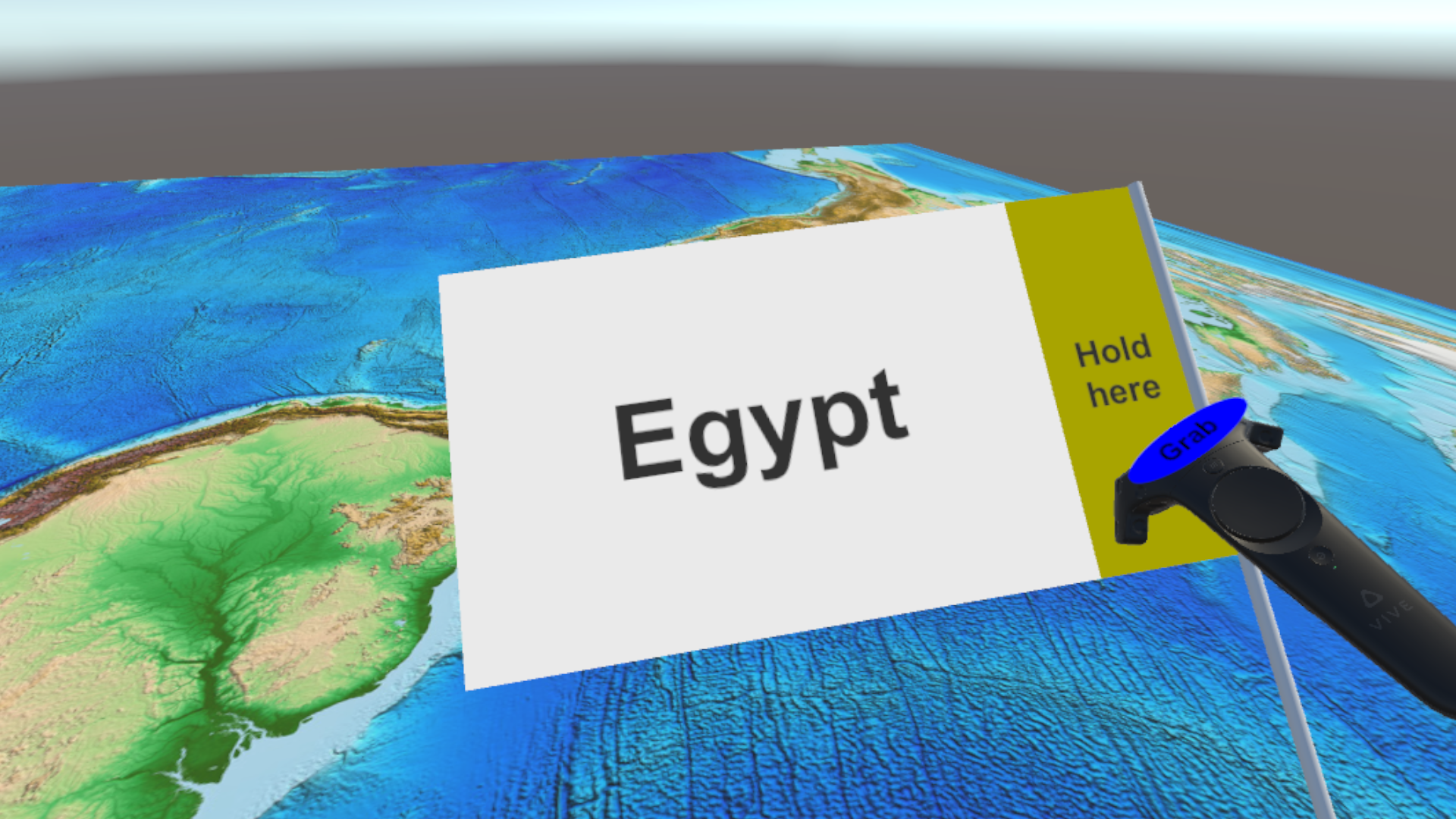}
\caption{Grabbable sticker.}
\label{fig:map-stickers}
\end{figure}

\subsection{Tutorial}
The experiment starts with the tutorial scenario, where the participant familiarizes with all the necessary controls required in the following scenario. These interactions are navigation in the virtual environment via walking or by teleportation and grabbing the virtual object.

\subsection{Geovisualization}
\begin{figure}[!htbp]
  	\centering
  	\includegraphics[clip,width=3.3in]{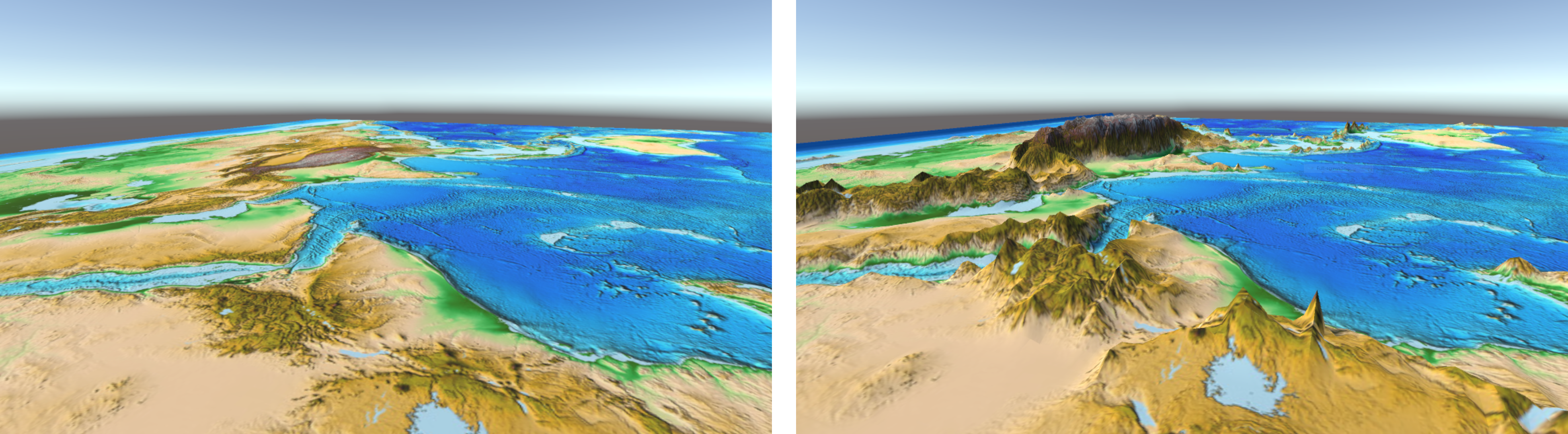}
	\caption{Map in 2D (left) and 3D (right) mode.}
	\label{fig:map-2d-3d}
\end{figure}

The scenario tests the usability of collaborative virtual environment in geovisualization with the goal to teach geography. Participants were instructed to grab stickers and put them on the map (Fig. \ref{fig:map-stickers}). The experimenter on top of that can change the environment. He can toggle between flat and 3D world map (Fig. \ref{fig:map-2d-3d}), change the type of map to visualize states and evaluate answers.


Participants in the role of a students performed the following task. The student places stickers with a geographical location on the map. There are four parts, where each part focuses on different geographical targets. These groups consist of states, islands, oceans and mountains. After putting stickers on the desired spot, the experimenter in the role of teacher evaluated results by pressing the appropriate button. Then, results were revealed to both the teacher and the student. After that, the teacher had an opportunity to give additional comments to the results and teach the student correct answers in case of a mistake.

\subsection{Participants and procedure}
The testing was performed in two separate rooms due to safety reasons. Voice communication was ensured by the Skype application. Thanks to the microphone built in the VR headset and connected headphones, the voice was loud and clear. Interaction itself happened within the virtual environment, where users could see each other's avatars. 

Participants were wearing the VR headset and the pair of controllers to interact in the virtual environment. They could move freely in an area of four by four meters.

We have conducted the user study with 30 participants consisting of 15 males and 15 females in the age between 18 to 33. All use computers on a daily basis. Some of them had a little experience with VR, mostly in a passive way without too much of an interaction.

\subsection{Data Collection}
Qualitative data were gathered from each participant after the experiment. Questions topics consisted of the reactions of the system, ease of learning, interface capabilities and open comments.

\section{RESULTS}
In this section, we summarize remarks of the user's behavior and interactions, that the experimenter noticed during the experiment, and also feedback from participants.

\subsection{Observations}
Most of the participants enjoyed the experience, especially if this was their first try of the immersive virtual environment. They mostly learned the controls for the first time without the need of repeated explaining. They quickly adapted to the movement with the teleportation. The participants were usually amazed, when the map of the world switched from 2D to 3D (Fig. \ref{fig:map-2d-3d}). It was also possible to read other user's body language to a little extent, thanks to the visibility of controllers and head orientation. When the participants were thinking of their answers, they tended to cross their hands (controllers swapped from left to right and vice versa and folded to the ``body'') and they kept staying on one place without teleportation. This might be an interesting topic of some future research and development of this application. Adding at least upper humanoid body might help to visualize body language more precisely.

\subsection{Feedback}
Participants responded in positive ways. One of the typical answers is as follows: ``Very intuitive controls, at least for someone who has experience with videogames, but has never experienced VR before. Items, and interactivity in general, are well done, manipulating with items seems natural. Overall this is a wonderful application held back by the hardware uses. Display resolution and mostly cables somewhat disturb the immersion''. Some participants pointed out, that it took them a while to figure out, that they can use the two controllers separately for two different interactions, e.g. grabbing with one hand and teleporting with the other hand. Here is one of the answers mentioning this issue: ``It took me some time to figure out that I can use my left hand as well -- with both hands it was much easier and more fun. At the beginning, it was difficult to switch the functions.'' On the other hand, they liked the teleporting feature, which made the experience more game-like.

\section{CONCLUSION AND FUTURE WORK}
This paper presented a generic collaborative virtual environment. A tutorial and a geovisualization scenario was implemented to evaluate its feasibility. 30 participants took part into the experiment and results were used to confirm the usability of the generic collaborative virtual environment in teaching geography. The follow-up research could explore additional methods of learning by using immersive collaborative virtual environment. It could also focus on differences in using simplified avatars compared to humanoid avatars. Finally, the shared VR environment will be used for cultural heritage.




\section*{ACKNOWLEDGMENT}
Part of this research was sponsored by the project ``Influence of cartographic visualization methods on the success of solving practical and educational spatial tasks'' (MUNI/M/0846/2015) and i-MareCulture project ``Advanced
VR, iMmersive Serious Games and Augmented REality as
Tools to Raise Awareness and Access to European Underwater
CULTURal heritagE, Digital Heritage'' (727153).



\bibliographystyle{IEEEtran}
\bibliography{IEEEabrv,bibliography.bib}

\begin{thebibliography}{10}
\providecommand{\url}[1]{#1}
\csname url@samestyle\endcsname
\providecommand{\newblock}{\relax}
\providecommand{\bibinfo}[2]{#2}
\providecommand{\BIBentrySTDinterwordspacing}{\spaceskip=0pt\relax}
\providecommand{\BIBentryALTinterwordstretchfactor}{4}
\providecommand{\BIBentryALTinterwordspacing}{\spaceskip=\fontdimen2\font plus
\BIBentryALTinterwordstretchfactor\fontdimen3\font minus
  \fontdimen4\font\relax}
\providecommand{\BIBforeignlanguage}[2]{{%
\expandafter\ifx\csname l@#1\endcsname\relax
\typeout{** WARNING: IEEEtran.bst: No hyphenation pattern has been}%
\typeout{** loaded for the language `#1'. Using the pattern for}%
\typeout{** the default language instead.}%
\else
\language=\csname l@#1\endcsname
\fi
#2}}
\providecommand{\BIBdecl}{\relax}
\BIBdecl

\bibitem{Feng2007}
\BIBentryALTinterwordspacing
W.-c. Feng, D.~Brandt, and D.~Saha, ``A long-term study of a popular mmorpg,''
  in \emph{Proceedings of the 6th ACM SIGCOMM Workshop on Network and System
  Support for Games}, ser. NetGames '07.\hskip 1em plus 0.5em minus 0.4em\relax
  New York, NY, USA: ACM, 2007, pp. 19--24. [Online]. Available:
  \url{http://doi.acm.org/10.1145/1326257.1326261}
\BIBentrySTDinterwordspacing

\bibitem{redfern_naughton_2002}
\BIBentryALTinterwordspacing
S.~Redfern and N.~Naughton, ``Collaborative virtual environments to support
  communication and community in internet-based distance education,''
  \emph{Journal of Information Technology Education}, vol.~1, no.~3, p.
  201–211, 2002. [Online]. Available:
  \url{https://aran.library.nuigalway.ie/handle/10379/4076}
\BIBentrySTDinterwordspacing

\bibitem{Benford2001}
\BIBentryALTinterwordspacing
S.~Benford, C.~Greenhalgh, T.~Rodden, and J.~Pycock, ``Collaborative virtual
  environments,'' \emph{Commun. ACM}, vol.~44, no.~7, pp. 79--85, Jul. 2001.
  [Online]. Available: \url{http://doi.acm.org/10.1145/379300.379322}
\BIBentrySTDinterwordspacing

\bibitem{McCloyRobert2001}
\BIBentryALTinterwordspacing
R.~McCloy and R.~Stone, ``\BIBforeignlanguage{English}{Virtual reality in
  surgery},'' \emph{\BIBforeignlanguage{English}{BMJ : British Medical
  Journal}}, vol. 323, no. 7318, p. 912, Oct 20 2001, copyright - Copyright:
  2001 (c) 2001 BMJ Publishing Group Ltd; Poslední aktualizace - 2016-04-04.
  [Online]. Available:
  \url{https://search.proquest.com/docview/1778069161?accountid=16531}
\BIBentrySTDinterwordspacing

\bibitem{Rizzo2011}
\BIBentryALTinterwordspacing
A.~Rizzo, T.~D. Parsons, B.~Lange, P.~Kenny, J.~G. Buckwalter, B.~Rothbaum,
  J.~Difede, J.~Frazier, B.~Newman, J.~Williams, and G.~Reger, ``Virtual
  reality goes to war: A brief review of the future of military behavioral
  healthcare,'' \emph{Journal of Clinical Psychology in Medical Settings},
  vol.~18, no.~2, pp. 176--187, 2011. [Online]. Available:
  \url{http://dx.doi.org/10.1007/s10880-011-9247-2}
\BIBentrySTDinterwordspacing

\bibitem{Faiola2013}
\BIBentryALTinterwordspacing
A.~Faiola, C.~Newlon, M.~Pfaff, and O.~Smyslova, ``Correlating the effects of
  flow and telepresence in virtual worlds: Enhancing our understanding of user
  behavior in game-based learning,'' \emph{Computers in Human Behavior},
  vol.~29, no.~3, pp. 1113 -- 1121, 2013. [Online]. Available:
  \url{http://www.sciencedirect.com/science/article/pii/S0747563212002713}
\BIBentrySTDinterwordspacing

\bibitem{Nikiforos2013}
\BIBentryALTinterwordspacing
N.~M. Papachristos, I.~Vrellis, A.~Natsis, and T.~A. Mikropoulos, ``The role of
  environment design in an educational multi-user virtual environment,''
  \emph{British Journal of Educational Technology}, vol.~45, no.~4, pp.
  636--646, 2014. [Online]. Available:
  \url{http://dx.doi.org/10.1111/bjet.12056}
\BIBentrySTDinterwordspacing

\bibitem{Pan200620}
\BIBentryALTinterwordspacing
Z.~Pan, A.~D. Cheok, H.~Yang, J.~Zhu, and J.~Shi, ``Virtual reality and mixed
  reality for virtual learning environments,'' \emph{Computers \& Graphics},
  vol.~30, no.~1, pp. 20 -- 28, 2006. [Online]. Available:
  \url{http://www.sciencedirect.com/science/article/pii/S0097849305002025}
\BIBentrySTDinterwordspacing

\bibitem{Garau2003}
\BIBentryALTinterwordspacing
M.~Garau, M.~Slater, V.~Vinayagamoorthy, A.~Brogni, A.~Steed, and M.~A. Sasse,
  ``The impact of avatar realism and eye gaze control on perceived quality of
  communication in a shared immersive virtual environment,'' in
  \emph{Proceedings of the SIGCHI Conference on Human Factors in Computing
  Systems}, ser. CHI '03.\hskip 1em plus 0.5em minus 0.4em\relax New York, NY,
  USA: ACM, 2003, pp. 529--536. [Online]. Available:
  \url{http://doi.acm.org/10.1145/642611.642703}
\BIBentrySTDinterwordspacing

\bibitem{billinghurst_kato1999}
M.~Billinghurst and H.~Kato, ``Collaborative mixed reality,'' in
  \emph{Proceedings of the First International Symposium on Mixed Reality},
  ser. ISMR’99, 1999, pp. 261--284.

\bibitem{Billinghurst2002}
\BIBentryALTinterwordspacing
------, ``Collaborative augmented reality,'' \emph{Commun. ACM}, vol.~45,
  no.~7, pp. 64--70, Jul. 2002. [Online]. Available:
  \url{http://doi.acm.org/10.1145/514236.514265}
\BIBentrySTDinterwordspacing

\bibitem{Potel1998}
M.~Potel, ``Motion sick in cyberspace,'' \emph{IEEE Computer Graphics and
  Applications}, vol.~18, no.~1, pp. 16--21, Jan 1998.

\bibitem{Bozgeyikli2016}
\BIBentryALTinterwordspacing
E.~Bozgeyikli, A.~Raij, S.~Katkoori, and R.~Dubey, ``Point \&\#38; teleport
  locomotion technique for virtual reality,'' in \emph{Proceedings of the 2016
  Annual Symposium on Computer-Human Interaction in Play}, ser. CHI PLAY
  '16.\hskip 1em plus 0.5em minus 0.4em\relax New York, NY, USA: ACM, 2016, pp.
  205--216. [Online]. Available:
  \url{http://doi.acm.org/10.1145/2967934.2968105}
\BIBentrySTDinterwordspacing

\bibitem{tambovtsev_floksy_peshe_2016}
\BIBentryALTinterwordspacing
D.~Tambovtsev, N.~Floksy, and O.~Peshé, ``How to avoid the effect of motion
  sickness in vr,'' Jul. 2016. [Online]. Available:
  \url{http://vrscout.com/news/avoid-motion-sickness-developing-for-vr/}
\BIBentrySTDinterwordspacing

\bibitem{Tregillus2016}
\BIBentryALTinterwordspacing
S.~Tregillus and E.~Folmer, ``Vr-step: Walking-in-place using inertial sensing
  for hands free navigation in mobile vr environments,'' in \emph{Proceedings
  of the 2016 CHI Conference on Human Factors in Computing Systems}, ser. CHI
  '16.\hskip 1em plus 0.5em minus 0.4em\relax New York, NY, USA: ACM, 2016, pp.
  1250--1255. [Online]. Available:
  \url{http://doi.acm.org/10.1145/2858036.2858084}
\BIBentrySTDinterwordspacing

\end{thebibliography}


\end{document}